\documentclass[10pt,twocolumn,letterpaper]{article}

\usepackage{cvpr}              

\usepackage{multirow}
\usepackage{makecell}
\usepackage{colortbl}
\usepackage[table]{xcolor}
\definecolor{tablegray}{gray}{0.9}
\usepackage{bm}
\usepackage{amsmath}
\usepackage{amssymb}
\usepackage{algorithm}
\usepackage{algpseudocode}
\usepackage{etoc}

\algnewcommand\algorithmicforeach{\textbf{for each}}
\algdef{S}[FOR]{ForEach}[1]{\algorithmicforeach\ #1\ \algorithmicdo}
\usepackage[most]{tcolorbox}
\tcbuselibrary{breakable, skins}
\usepackage{listings}
\usepackage{enumitem}
\newtcolorbox{promptbox}[1][]{
  colback=gray!5,
  colframe=gray!40,
  title=\textbf{Prompt Content},
  fonttitle=\bfseries\sffamily,
  fontupper=\sffamily\small,
  arc=0mm,
  boxrule=0.5pt,
  enhanced,
  breakable,
  #1
}
\tcbuselibrary{breakable, skins}
\newtcolorbox{reasoningbox}[2][]{
  colback=gray!5,       
  colframe=gray!40,     
  coltitle=black,       
  title=\textbf{#2},
  fonttitle=\bfseries\sffamily,
  fontupper=\small,
  boxrule=0.8pt,
  arc=3mm,              
  enhanced,
  breakable,            
  #1
}

\definecolor{cvprblue}{rgb}{0.21,0.49,0.74}
\usepackage[pagebackref,breaklinks,colorlinks,allcolors=cvprblue]{hyperref}

\def\paperID{} 
\def\confName{CVPR}
\def\confYear{2026}

\title{Pixel2Phys: Distilling Governing Laws from Visual Dynamics}

\author{
Ruikun Li$^{2}$\thanks{Work done during an internship at Shanghai AI Laboratory.}~\thanks{Equal contribution.}, 
Jun Yao$^{1,3}$\footnotemark[2], 
Yingfan Hua$^{1}$, 
Shixiang Tang$^{1,4}$, 
Biqing Qi$^1$, \\
Bin Liu$^{3}$, 
Wanli Ouyang$^{1,4}$, 
Yan Lu$^{1,4}$\thanks{Corresponding author (luyan@pjlab.org.cn)}\\
$^1$Shanghai Artificial Intelligence Laboratory
$^2$Tsinghua University \\
$^3$University of Science and Technology of China 
$^4$The Chinese University of Hong Kong\\
}

\begin{document}
\etocdepthtag.toc{main}

\maketitle
\begin{abstract}
Discovering physical laws directly from high-dimensional visual data is a long-standing human pursuit but remains a formidable challenge for machines, representing a fundamental goal of scientific intelligence. 
This task is inherently difficult because physical knowledge is low-dimensional and structured, whereas raw video observations are high-dimensional and redundant, with most pixels carrying little or no physical meaning. 
Extracting concise, physically relevant variables from such noisy data remains a key obstacle. 
To address this, we propose Pixel2Phys, a collaborative multi-agent framework adaptable to any Multimodal Large Language Model (MLLM). 
It emulates human scientific reasoning by employing a structured workflow to extract formalized physical knowledge through iterative hypothesis generation, validation, and refinement.
By repeatedly formulating, and refining candidate equations on high-dimensional data, it identifies the most concise representations that best capture the underlying physical evolution. 
This automated exploration mimics the iterative workflow of human scientists, enabling AI to reveal interpretable governing equations directly from raw observations. 
Across diverse simulated and real-world physics videos, Pixel2Phys discovers accurate, interpretable governing equations and maintaining stable long-term extrapolation where baselines rapidly diverge.
\end{abstract}    
\begin{figure*}
    \centering
    \includegraphics[width=1\linewidth]{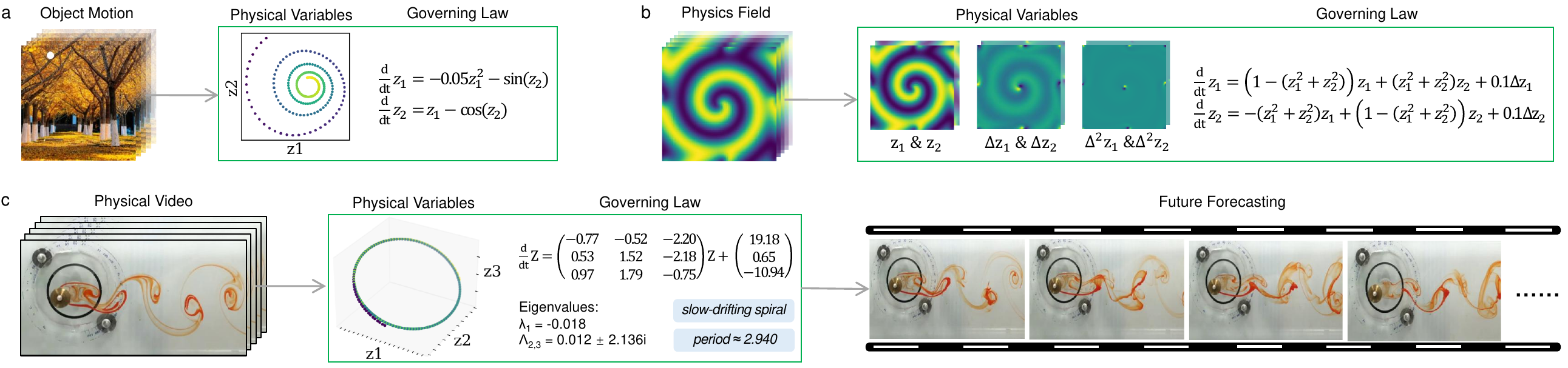}
    \caption{
    Distill governing laws from videos. 
    (a) Trajectory dynamics of moving objects.
    (b) Spatiotemporal dynamics of time-varying physical fields.
    (c) Intrinsic dynamics of physical phenomena.
    }
    \label{fig:intro}
    \vspace{-0.3cm}
\end{figure*}

\section{Introduction}
Discovering physical laws from observational data lies at the core of scientific understanding and has historically driven major advances across physics, astronomy, and the natural sciences~\cite{tenachi2023deep, wang2019symbolic, angelis2023artificial, cao2024teaching}.  
As AI systems increasingly participate in scientific workflows~\cite{ying2025neural, liu2024sora}, the ability to infer governing equations directly from visual observations becomes a critical step toward genuine AI for Science.  
Human scientists typically achieve such discoveries by distilling structured, low-dimensional physical variables from complex visual phenomena and formulating compact symbolic laws~\cite{pope2021intrinsic, chen2022automated}.  
However, this manual process is slow and labor-intensive, classic examples such as translating Tycho Brahe’s astronomical measurements into Kepler’s laws required decades of expert reasoning.  
Automating this capability, \emph{visual equation discovery}, would therefore accelerate scientific progress.

However, visual equation discovery is extremely challenging because the true physical signals in a video are low-dimensional and concise, yet they are submerged within a large amount of visually irrelevant content such as textures, lighting variations, and background motion~\cite{}.  
These redundant components dominate the pixel space and obscure the compact physical structure that scientific laws depend on.
Prior approaches have attempted to address this difficulty in several ways, but each faces fundamental limitations.  
(1) Supervised equation–prediction models~\cite{biggio2021neural, shojaee2023transformer, la2021contemporary, ying2025neural}: learn to directly predict equation from videos, but they require manually organzied equation–video data pairs for training, which are extremely scarce, causing the models to generalize poorly beyond the specific systems they were trained on.  
(2) Unsupervised latent-coding methods: first learn latent representations from videos using autoencoding or predictive objectives~\cite{guen2020disentangling, wu2023disentangling, higgins2021symetric,hofherr2023neural, bounou2021online,li2023learning}, and then apply symbolic regression on these latents.  
However, because the latent spaces are determined by reconstruction or next-frame prediction optimization process rather than physical structure, they are often non-unique and easily entangle visually salient but physically irrelevant factors. 
(3) Recent Multimodal Large Language Models (MLLMs) combine visual understanding with strong symbolic reasoning, suggesting potential for equation discovery~\cite{boiko2023autonomous,birhane2023science,li2025mllm}.  
Yet directly prompting an MLLM mainly retrieves and recombines prior knowledge from its training corpus; without a structured workflow, it struggles to infer new physical variables or derive laws purely from raw visual data.
These challenges highlight the need for a method that can reliably extract the physically meaningful structure hidden inside high-dimensional videos and support the discovery of new equations.

To address these challenges, we propose Pixel2Phys, a collaborative multi-agent framework that can be paired with any MLLM and organizes visual equation discovery into a structured, iterative scientific workflow (Figure~\ref{fig:framework}a).  
At the center of the framework is the Plan Agent, which coordinates three specialized agents, the Variable Agent, the Equation Agent, and the Experiment Agent, and determines the refinement strategy across iterations.
The Variable Agent extracts physical quantities from videos through several complementary defined tools that correspond to different classes of physical systems, enabling it to recover object-level motion, field-level evolution, and other low-dimensional structures that often underlie governing equations.  
The Equation Agent forms equation candidates by dynamically utilizing symbolic regression components, such as adjusting sparsity constraint strength or selecting different symbolic libraries.  
The Experiment Agent evaluates each candidate equation by several metrics, such as simulating its predicted dynamics, measuring reconstruction discrepancies, testing temporal extrapolation, and summarizing these results into a structured report for the Plan Agent.
The Plan Agent integrates these reports and issues targeted instructions, such as enforcing stronger sparsity in the Equation Agent or requiring the Variable Agent to incorporate dynamical constraints, for a next refinement step.  
Through this iterative cycle of hypothesis formation, evaluation, and adjustment, Pixel2Phys progressively filters away visually irrelevant factors and converges toward variables and equations that best describe the true underlying dynamics.
Crucially, each step in this workflow is mechanical and fully within the execution capabilities of modern MLLMs, allowing Pixel2Phys to combine the model’s strong single-step reasoning with structured multi-step coordination, and thereby to discover new variables and governing equations beyond what is present in the model’s training data.  
We evaluate our framework on three categories of physics videos with increasing difficulty. The results demonstrate that by distilling accurate and interpretable governing equations, our proposed framework improves extrapolation accuracy (RMSE) by 45.35\% over baselines.

Our contributions can be summarized as follows:
\begin{itemize}
    \item We propose a novel multi-agent framework for visual equation discovery, where an MLLM planner coordinates specialized agents to parse complex visual dynamics at multiple granularities.
    \item We design an iterative, co-optimization reasoning process that breaks the circular dependency between visual representation learning and law discovery.
    \item Extensive experiments on three challenging categories of physics videos demonstrate that our framework not only discovers physically interpretable governing equations but also improves long-term extrapolation accuracy by 45.35\%.
\end{itemize}

\section{Related Work}

\subsection{Inferring Physics from Video}
Existing approaches generally tackle visual dynamics from two perspectives. 
The first category focuses on \textbf{implicit neural dynamics}, learning latent representations to perform future prediction. 
Modules like Neural ODEs~\cite{hofherr2023neural, chen2018neural} and Koopman operators~\cite{bounou2021online} are often integrated to model continuous evolution. 
Other works decompose video into PDE dynamics~\cite{guen2020disentangling, wu2023disentangling} or disentangled representations~\cite{yang2022learning, fotiadis2023disentangled}. 
While effective for prediction, these methods encapsulate physical laws within black-box networks, lacking explicit interpretability.
The second category, \textbf{physics-informed perception}, imposes strong inductive biases. 
Researchers have utilized inverse graphics~\cite{jaques2019physics} or enforced rigid body constraints~\cite{kandukuri2022physical} to infer specific properties like mass and friction. 
However, these methods rely heavily on \textit{a priori} knowledge of the governing equations, limiting their ability to discover unknown laws from unfamiliar physical phenomena.


\subsection{Visual Equation Discovery}
Equation discovery from video is a nascent field that typically follows two paradigms: 
(1) \textbf{Pipeline-based approaches}~\cite{luan2022distilling, zhang2024vision}, which extract explicit trajectories of objects and then apply symbolic regression. These methods rely on pre-trained trackers and struggle with continuous fields (e.g., fluids) where objects are undefined.
(2) \textbf{End-to-end approaches}~\cite{udrescu2021symbolic, champion2019data}, which learn a latent coordinate space for equation regression. However, they face a critical \textbf{circular dependency}: extracting a good variable space requires knowledge of the dynamics, while finding the dynamics requires a clean variable space. 
Consequently, they often settle for sub-optimal solutions in complex visual scenarios.
\textbf{Our work introduces a third paradigm:} a collaborative, multi-agent framework. By establishing a reasoning-driven feedback loop, we enable variable extraction and equation discovery to mutually refine each other, effectively breaking the circular dependency and extending discovery to diverse visual dynamics.

\section{Problem Fomulation}~\label{sec:problem}
In the task of inferring governing laws from high-dimensional data, the goal is to find a compact and accurate symbolic expression.
We are given a sequence of high-dimensional visual observations (i.e., a video), $\{\mathbf{X}(t)\}_{t=1}^N$, where each frame $\mathbf{X}(t) \in \mathbb{R}^D$ ($D$ is the pixel space, e.g., $C \times H \times W$). We assume this video observes an underlying, unknown, low-dimensional dynamical system evolving on an intrinsic manifold~\cite{champion2019data}, $\mathcal{Z} \subseteq \mathbb{R}^d$, where $d \ll D$.
Our goal is to simultaneously discover both the intrinsic physical variables $z(t) \in \mathcal{Z}$ and their symbolic governing law $f$, which defines the system's evolution, $\frac{dz}{dt} = f(z(t))$.
This presents a challenging dual discovery problem, as both the coordinate system $z$ and the function $f$ are unknown. This problem can be decomposed into two coupled sub-problems:
\begin{itemize}[leftmargin=*]
    \item \textbf{Physical Variable Extraction.} 
    Learning an encoder $\phi: \mathbb{R}^D \rightarrow \mathbb{R}^d$ that maps the high-dimensional visual observation $\mathbf{X}(t)$ to its corresponding intrinsic physical state $z(t) = \phi(\mathbf{X}(t))$.
    To ensure $z(t)$ is an informative representation, this component often includes a decoder $\psi: \mathbb{R}^d \rightarrow \mathbb{R}^D$, used to enforce a reconstruction constraint $\mathbf{X}(t) \approx \psi(z(t))$.
    
    \item \textbf{Governing Law Distillation.} 
    Identifying an interpretable, symbolic expression for the dynamics function $f$ from a library of candidate functions $\Theta$ (e.g., polynomials, trigonometric terms). This function must accurately model the evolution of the extracted variables $z(t)$.
\end{itemize}
The synergy between these components is the central challenge. The quality of the extracted variables $z = \phi(\mathbf{X})$ directly dictates the simplicity and accuracy of the distilled law $f$. Conversely, a simple and sparse law $f$ provides a powerful signal to guide the variable extraction process $\phi$, compelling it to filter out dynamically irrelevant visual components. 
Our framework is designed to solve this co-optimization problem, seeking a self-consistent pair of an intrinsic variable space $\mathcal{Z}$ and its symbolic law $f$ that fit the observed data and generalize for long-term prediction.

\begin{figure*}[ht]
    \centering
    \includegraphics[width=0.97\linewidth]{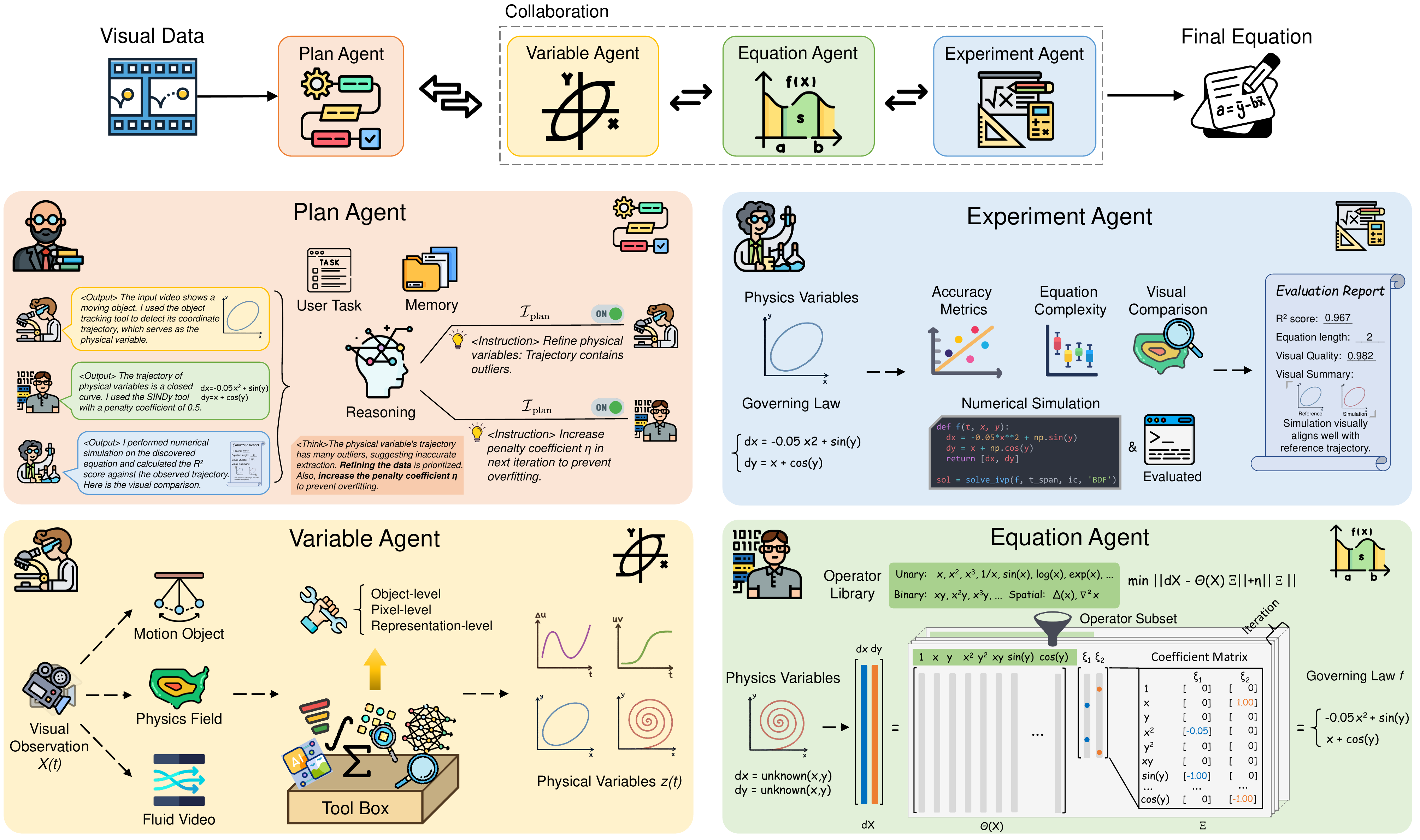}
    \caption{The multi-agent collaboration framework of Pixel2Phys.}
    \label{fig:framework}
    \vspace{-0.3cm}
\end{figure*}

\section{Method}

\subsection{Pixel2Phys Method Overview}
The Pixel2Phys framework mimics the collaborative workflow of a human scientific team to solve the dual discovery problem defined in Section~\ref{sec:problem}, involving observing, hypothesizing, experimenting, and refining.
As illustrated in Figure~\ref{fig:framework}, Pixel2Phys consists of four agents with distinct roles: 
The Plan Agent acts as the team's central coordinator, responsible for setting goals, analyzing reports, diagnosing bottlenecks, and providing instructional prompts $\mathcal{I}_{\text{plan}}$ to other agents to guide the next iteration. 
The Variable Agent executes visual parsing tasks, i.e., parsing and extracting low-dimensional physical variables $z(t)$ from the high-dimensional observations $\mathbf{X}(t)$. 
The Equation Agent is responsible for distilling the symbolic governing equation $f$ from $z(t)$. 
Finally, the Experiment Agent validates the quality of the current $(\mathcal{Z}, f)$ pair.
In each iteration, the Variable, Equation, and Experiment Agent are required to return a report  (denoted as $\mathcal{R}_{var}$, $\mathcal{R}_{equ}$, and $\mathcal{R}_{exp}$) to the Plan Agent, respectively, containing both quantitative performance metrics and qualitative visual diagnostics for the next decision-making step.
These agents are instantiated as MLLMs and operate under a unified protocol where they process visual data $\mathbf{X}$ and textual prompts $\mathcal{P}$ to generate textual responses $\text{MLLM}(\mathbf{X}, \mathcal{P})$. 
Algorithm~\ref{app:pseudocode} outlines the pseudocode of Pixel2Phys's execution process.

\subsection{Plan Agent: Global Planning in Iterative Reasoning}
Our framework replaces a simple pipeline with an iterative reasoning loop driven by the Plan Agent. 
This agent is the central coordinator that orchestrates the co-optimization of the variable space $\mathcal{Z}$ and the governing law $f$.

The reasoning loop commences with an initialization step ($k=0$).
The Plan Agent interprets the user's task and activates the initial iteration, yielding the candidate pair $(\mathcal{Z}_0, f_0)$ alongside the reports $\mathcal{R}_{\text{var}}^{0}$, $\mathcal{R}_{\text{equ}}^{0}$, and $\mathcal{R}_{\text{exp}}^{0}$ from each agent.
In subsequent iterations ($k \ge 1$), the Plan Agent aggregates these reports to conduct a two-fold diagnosis. 
It first inspects the visualizations in $\mathcal{R}_{\text{exp}}^{k-1}$ to assess qualitative dynamical fidelity, then scrutinizes the specific quantitative metrics with $\mathcal{R}_{\text{var}}^{k-1}$ and $\mathcal{R}_{\text{equ}}^{k-1}$ to pinpoint the exact bottleneck. 
Based on this analysis, it formulates a instructional prompt $\mathcal{I}_{\text{plan}}^k$ to resolve the identified failure mode:
\begin{itemize}
    \item \textbf{Variable Refinement.}
    When the diagnosis attributes failure to a poor $\mathcal{Z}$ (e.g., high reconstruction errors), the Plan Agent instructs the Variable Agent to re-extract $\mathcal{Z}$. Crucially, Plan Agent will provide the equation $f_{k-1}$ to activate the physics-informed loss (Section~\ref{sec:physics_informed}).
    
    \item \textbf{Equation Refinement.}
    When $\mathcal{Z}$ is deemed high-quality but $f_{k-1}$ is inaccurate or overly complex, the Plan Agent instructs the Equation Agent to modify its equation search by adjusting the configuration hyperparameters.
\end{itemize}
This iterative refinement loop continues until the Plan Agent determines that $\mathcal{R}_{\text{exp}}^k$ satisfies the success criteria (Appendix~\ref{app:our_config}). 
The final $(\mathcal{Z}_k, f_k)$ pair is then returned as the solution.

\subsection{Variable Agent:  Variable Extraction from Visual Data}
Given a video sequence $\mathbf{X}(t)$, the Variable Agent extracts physical variables $z(t)$ by deploying specific parsing tools. 
We provide multi-granularity tools for flexible usage to accommodate physical information presented at object, pixel, and representation levels.
The agent dynamically selects the corresponding tool based on both the video's visual properties and the explicit instructional prompt $\mathcal{I}_{\text{plan}}$ issued by the Plan Agent.
Specifically, we provide object-level tools for videos of moving objects, pixel-level tools for physical fields, and autoencoder-based tools for complex scientific phenomena.

\paragraph{Object-level Tool.}
For moving objects, such as celestial revolution, the Variable Agent extracts the trajectory $z(t) = [x(t), y(t)]$. 
We employ visual foundation models for zero-shot object segmentation and tracking to avoid training on specific shapes of objects. 
Specifically, we utilize Segment Anything~\cite{kirillov2023segment} to segment potential objects in every frame. 
The agent then filters out static targets, as detailed in Appendix~\ref{sec:appendix_static}, and records the centroid coordinates of moving objects as $z(t)$.

\paragraph{Pixel-level Tool.}
For physical fields governed by PDEs (Figure~\ref{fig:intro}b), video frames $\mathbf{X}(t)$ are treated as samples of a continuous field $u(\mathbf{x}, t)$ discretized at pixels $\mathbf{x}$~\cite{cao2024teaching, wu2024pastnet}. 
Physical dynamics emerge from local spatial interactions. 
We equip the Variable Agent with the fixed convolutional kernels to compute spatial derivatives directly from the pixel grid, yielding operators such as the Laplacian $\Delta$ and biharmonic $\Delta^2$.
Since the governing equation is valid at any pixel, the agent randomly samples a sparse subset of pixels to efficiently collect the spatial operators as physical variables $z(t) = [u(t), \Delta u(t), \Delta^2 u(t), \dots]$.

\paragraph{Representation-level Tool.}\label{sec:physics_informed}
For complex scientific phenomena, underlying dynamics are often obscured by device noise and lighting fluctuations (Figure~\ref{fig:intro}c). 
We adapt a novel physics-informed autoencoder to compress $\mathbf{X}(t)$ into a latent representation $z(t) = \phi(\mathbf{X}(t))$, ensuring the underlying variables remain as simple as possible. 
This tool operates in two modes depending on the availability of physical priors.
First, when the instructional prompt $\mathcal{I}_{\text{plan}}$ contains a governing equation $f$ discovered in the previous iteration, the loss function comprises two components $
\mathcal{L} = \mathcal{L}_{\text{recon}} + \lambda_{\text{eq}} \mathcal{L}_{\text{eq}}$.
The first term is the self-supervised reconstruction error ensuring $z$ retains sufficient observational information. The second term represents the physics consistency loss with coefficient $\lambda_{\text{eq}}$ defined as $\mathcal{L}_{\text{eq}} = \| \mathcal{F}(z) - f(z) \|^2$,
where $\mathcal{F}(z)$ denotes the numerical derivative via central differences and $f(z)$ represents the symbolic derivative derived from equation $f$. 
This constraint forces the latent space to adhere to the governing equation $f$, guiding the encoder to focus on physical dynamics while filtering out textural details and noise.
Second, in the absence of physical information, such as during the cold-start phase, the autoencoder performs only self-supervised reconstruction and functions as a standard autoencoder. 
This approach ensures the Variable Agent identifies a latent space $\mathcal{Z}$ that balances visual reconstruction with dynamic simplicity. Crucially, it enables the output from the downstream Equation Agent to iteratively refine the extraction of physical variables.

Finally, the process concludes by returning the extracted variables $z(t)$ alongside a report $\mathcal{R}_{\text{var}}$. 
This report encapsulates the specific tools employed and their hyperparameter configurations, such as the model size of SAM and its mask size.

\subsection{Equation Agent: Dynamic Symbolic Regression}
Considering that most true equations $\frac{dz}{dt} = f(z(t))$ exhibit sparsity within a high-dimensional space of candidate functions~\cite{brunton2016discovering,champion2019data}, the agent identifies the sparse active terms in this function space through a three-step process.

\paragraph{Data and Derivative Estimation.}
Given the discrete time series $z(t)$ organized into a state matrix $Z$, the agent first estimates the time derivative $\dot{Z}$ numerically. 
This step utilizes the central difference method~\cite{swanson1992central} $\mathcal{F}(z)$ to ensure methodological consistency, thereby establishing the left-hand side (LHS) of the target equation.

\paragraph{Candidate Library Construction.}
Subsequently, the agent constructs a candidate library matrix $\Theta(Z)$, wherein each column represents a potential nonlinear function of the state $z(t)$ that constitutes the right-hand side (RHS) of the governing equation. 
The candidate library incorporates the following component terms
\begin{itemize}[leftmargin=*]
\item Polynomial terms including $1, z, z^2, z_1 z_2, \dots$
\item Transcendental terms $\sin(z), \cos(z), exp(z), \dots$.
\end{itemize}
The equation agent iteratively optimizes the candidate library in the following process.

\paragraph{Sparse Regression and Law Distillation.}
The discovery task is formulated as solving the overdetermined linear system $\dot{Z} \approx \Theta(Z) \Xi$,
where $\dot{Z}$ denotes the derivative matrix, $\Theta(Z)$ represents the candidate library, and $\Xi$ is the unknown sparse coefficient matrix.
The optimization objective is formulated as $\|\dot{Z} - \Theta(Z) \Xi\|^2_2 + \lambda_{sp} \|\Xi\|_1$, wherein the second term enforces the sparsity of active terms.
To solve for $\Xi$, the agent employs the Sequential Thresholded Least-Squares (STLSQ) algorithm~\cite{brunton2016discovering} (detailed in Appendix~\ref{sec:appendix_STLSQ}). 
The sparsity threshold $\lambda_{sp}$ is determined via the Plan Agent's instruction $\mathcal{I}_{\text{plan}}$, enabling active guidance over the parsimony of the discovered law. 
Ultimately, the non-zero entries in $\Xi = [\xi_1, \xi_2, \dots, \xi_d]$ are reconstructed into symbolic equations $f$. 

Finally, the process concludes by returning the discovered equation $f$ and a report $\mathcal{R}_{\text{equ}}$, containing the candidate library and sparsity threshold $\lambda_{sp}$ for the decision of the Plan Agent.

\begin{figure*}[t]
    \centering
    \includegraphics[width=0.97\linewidth]{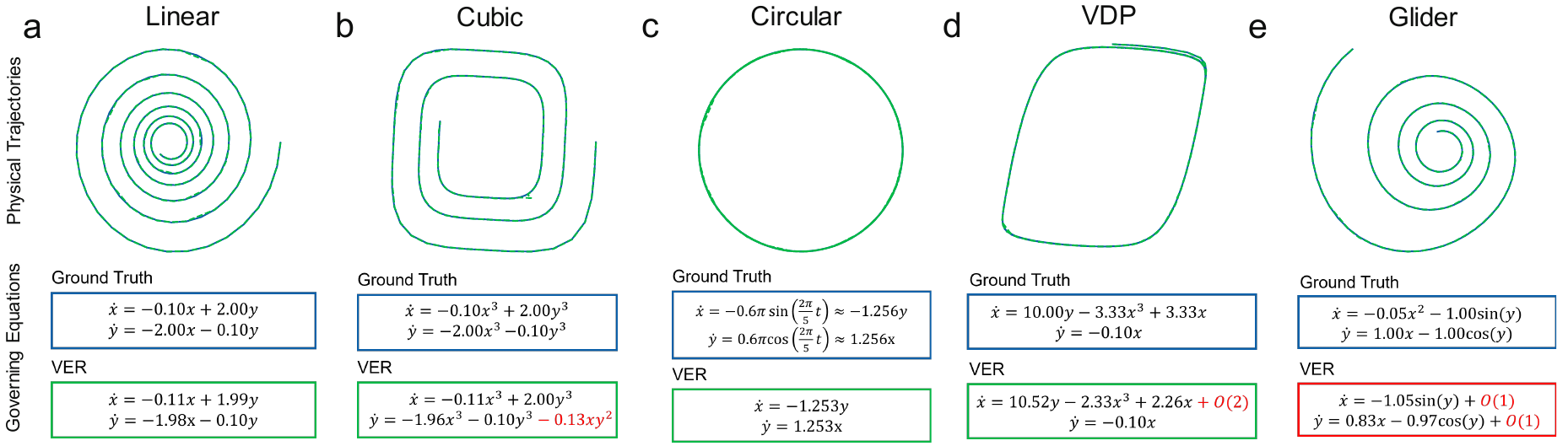}
    \caption{Reasoning results of motion objects: blue line is ground truth; green dashed line is trajectories inferred by PixelsPhys.}
    \label{fig:motion}
    \vspace{-0.2cm}
\end{figure*}

\subsection{Experiment Agent: Equation Evaluation and Feedback}
The Experiment Agent's role is to rigorously validate the quality and consistency of the $(\mathcal{Z}, f)$ pair discovered in the current iteration. 
It receives the variable time series $z(t)$ and the symbolic law $f$ and generates a multi-dimensional evaluation report, $\mathcal{R}_{\text{exp}}$, for the Plan Agent. 
This validation protocol assesses three key aspects.

\paragraph{Equation Quality.}
To assess the governing law $f$, Experiment Agent computes two quantitative metrics: (1) $R^2$ score by comparing the numerical derivative $\mathcal{F}(z)$ with the symbolic derivative $f(z)$; and (2) Complexity, measured by the number of terms, $L_0$, of the coefficient matrix $\Xi$.

\paragraph{Variable Quality.}
The Experiment Agent generates phase portraits of the trajectory $z(t)$ to visually characterize the variable space. Concise governing laws typically manifest as structured chaotic attractors, whereas noisy or tangled trajectories suggest the presence of significant interference.
This image is passed to the Plan Agent, which visually assesses whether the trajectory exhibits a clear, low-dimensional dynamical structure.

\paragraph{Extrapolation Fidelity.}
The agent performs a long-term numerical simulation $z_{\text{pred}}(t)$ by integrating the discovered law $f$ from an initial condition $z(0)$. 
It then (1) computes the Root Mean Square Error (RMSE) between the predicted $z_{\text{pred}}(t)$ and the extracted variables from held-out future frames $z_{\text{gt}}(t)$ over an unseen time horizon; and (2) visualizes the plots of the predicted and ground-truth trajectories.
A self-consistent pair $(\mathcal{Z}, f)$ yields reliable long-term predictions.

Finally, the Experiment Agent aggregates all these quantitative metrics ($R^2$, $L_0$, RMSE) and plotted figures into the structured report $\mathcal{R}_{\text{exp}}$ (see Appendix~\ref{sec:appendix_report} for structure), which serves as the foundation for the Plan Agent's next reasoning step.

\begin{table}[t]
\renewcommand{\arraystretch}{0.9}
\caption{Average performance of motion objects over 5 runs with varying seeds. Best results are in bold.}
\resizebox{\linewidth}{!}{%
\centering
\begin{tabular}{ccccc}
\toprule
 Case & Method & \makecell{Terms \\ Found} & \makecell{False \\ Positives} & $R^2@1000$  \\
\midrule
 \multirow{4}{*}{Linear} 
 & AE-SINDy & - & - & $0.0046_{\pm0.0028}$  \\
 & Latent-ODE & - & - & $0.0154_{\pm0.0081}$  \\
 \cmidrule(lr){2-5}
 & Coord-Equ & Yes & $1.100_{\pm0.3600}$ & $0.8647_{\pm0.0554}$  \\
 & Pixel2Phys & Yes & $\mathbf{0}$ & $\mathbf{0.9913_{\pm0.0000}}$  \\
\midrule
 \multirow{4}{*}{Cubic} 
 & AE-SINDy & - & - & $0.0720_{\pm0.0122}$  \\
 & Latent-ODE & - & - & $0.0039_{\pm0.0013}$  \\
 \cmidrule(lr){2-5}
 & Coord-Equ & No & $3.400_{\pm1.2800}$ & $0.2632_{\pm0.1928}$  \\
 & Pixel2Phys & Yes & $\mathbf{0.3900_{\pm0.3620}}$ & $\mathbf{0.9886_{\pm0.0082}}$  \\
\midrule
 \multirow{4}{*}{Circular} 
 & AE-SINDy & - & - & $0.3647_{\pm0.0716}$  \\
 & Latent-ODE & - & - & $0.0240_{\pm0.0028}$  \\
 \cmidrule(lr){2-5}
 & Coord-Equ & Yes & $0.2000_{\pm0.0040}$ & $0.9903_{\pm0.0057}$  \\
 & Pixel2Phys & Yes & \textbf{0} & $\mathbf{1.0000_{\pm0.0000}}$  \\
\midrule
 \multirow{4}{*}{VDP} 
 & AE-SINDy & - & - & $0.2483_{\pm0.0182}$  \\
 & Latent-ODE & - & - & $0.0433_{\pm0.0102}$  \\
 \cmidrule(lr){2-5}
 & Coord-Equ & Yes & $2.3100_{\pm0.6590}$ & $0.4920_{\pm0.1302}$  \\
 & Pixel2Phys & Yes & $\mathbf{0.9900_{\pm0.0030}}$ & $\mathbf{0.9954_{\pm0.0047}}$ \\
\midrule
 \multirow{4}{*}{Glider} 
 & AE-SINDy & - & - & $0.0310_{\pm0.0030}$  \\
 & Latent-ODE & - & - & $0.0360_{\pm0.0041}$  \\
 \cmidrule(lr){2-5}
 & Coord-Equ & No & $\mathbf{2.1800_{\pm0.4900}}$ & $0.9129_{\pm0.0102}$  \\
 & Pixel2Phys & No & $3.0000_{\pm0.0000}$ & $\mathbf{0.9995_{\pm0.0000}}$ \\
\bottomrule
\end{tabular}%
}%
\label{tab:motion}
\vspace{-0.3cm}
\end{table}

\section{Experiments}
We validate Pixel2Phys on three categories of interdisciplinary videos: object motion (object-level), physical fields (pixel-level), and real-world recordings of scientific phenomena (representation-level). We employ GPT-4o as LLM backbone by default. All configurations for Pixel2Phys across experiments are provided in Appendix~\ref{app:our_config}.

\begin{table*}[t]
\centering
\renewcommand{\arraystretch}{1}
\caption{Average RMSE and VPS (± std from 5 runs) of long-term prediction. Best results are in bold. The threshold of VPS is 0.5.}
\resizebox{\textwidth}{!}{%
\begin{tabular}{lcccccccc}
\toprule
 \multirow{2}{*}{Method} & \multicolumn{2}{c}{Lambda-Omega} & \multicolumn{2}{c}{Brusselator} & \multicolumn{2}{c}{ FitzHugh–Nagumo} & \multicolumn{2}{c}{Swift–Hohenberg} \\
 & \multicolumn{1}{c}{RMSE $\downarrow$} & \multicolumn{1}{c}{VPS@0.5 $\uparrow$} & \multicolumn{1}{c}{RMSE $\downarrow$} & \multicolumn{1}{c}{VPS@0.5 $\uparrow$} & \multicolumn{1}{c}{RMSE $\downarrow$} & \multicolumn{1}{c}{VPS@0.5 $\uparrow$} & \multicolumn{1}{c}{RMSE $\downarrow$} & \multicolumn{1}{c}{VPS@0.5 $\uparrow$} \\
 
 \midrule
 \multicolumn{9}{c}{\cellcolor{tablegray}Black-box Models} \\
 
 FNO~\cite{li2020fourier} & $0.68_{\pm 0.04}$ & $477.00_{\pm 20.82}$ & $415.34_{\pm 81.70}$ & $19.10_{\pm 3.96}$ & $0.89_{\pm 0.07}$ & $116.20_{\pm 6.81}$ & $11.02_{\pm 4.09}$ & $52.10_{\pm 9.04}$ \\
 
 UNO~\cite{rahman2022u} & $0.42_{\pm 0.05}$ & $764.80_{\pm 16.97}$ & $423.64_{\pm 82.42}$ & $27.30_{\pm 4.23}$ & $67.41_{\pm 5.26}$ & $104.00_{\pm 13.73}$ & $0.48_{\pm 0.08}$ & $90.80_{\pm 5.81}$ \\
 
 WNO~\cite{tripura2023wavelet} & $96.98_{\pm 6.35}$ & $41.20_{\pm 6.10}$ & $68.67_{\pm 7.80}$ & $9.60_{\pm 3.67}$ & $34.10_{\pm 3.09}$ & $22.60_{\pm 5.24}$ & $1.95_{\pm 0.39}$ & $31.30_{\pm 8.93}$ \\

 \multicolumn{9}{c}{\cellcolor{tablegray}Symbolic-regression Models} \\

 PDE-Find~\cite{rudy2017data} & $0.67_{\pm 0.00}$ & $492.00_{\pm 0.00}$ & $1.56_{\pm 0.00}$ & $40.00_{\pm 0.00}$ & $0.63_{\pm 0.00}$ & $54.00_{\pm 0.00}$ & $0.19_{\pm 0.00}$ & $200.00_{\pm 0.00}$ \\
 SGA-PDE~\cite{chen2022symbolic} & $0.92_{\pm 0.10}$ & $126.60_{\pm 6.51}$ & $0.14_{\pm 0.02}$ & $1000.00_{\pm 0.00}$ & NaN & NaN & NaN & NaN \\
 LLM-PDE~\cite{du2024large} & $0.55_{\pm 0.04}$ & $438.40_{\pm 23.20}$ & NaN & NaN & $0.62_{\pm 0.12}$ & $55.90_{\pm 4.20}$ & $0.69_{\pm 0.13}$ & $34.30_{\pm 13.03}$ \\
 
 Pixel2Phys & $\mathbf{0.03_{\pm 0.00}}$ & $\mathbf{1000.00_{\pm 0.00}}$ & $\mathbf{0.12_{\pm 0.01}}$ & $\mathbf{1000.00_{\pm 0.00}}$ & $\mathbf{0.16_{\pm 0.01}}$ & $\mathbf{1000.00_{\pm 0.00}}$ & $\mathbf{0.18_{\pm 0.06}}$ & $\mathbf{200.00_{\pm 0.00}}$ \\

\bottomrule
\end{tabular}%
}
\label{tab:pde}
\vspace{-0.1cm}
\end{table*}

\begin{figure*}[t]
    \centering
    \includegraphics[width=0.97\linewidth]{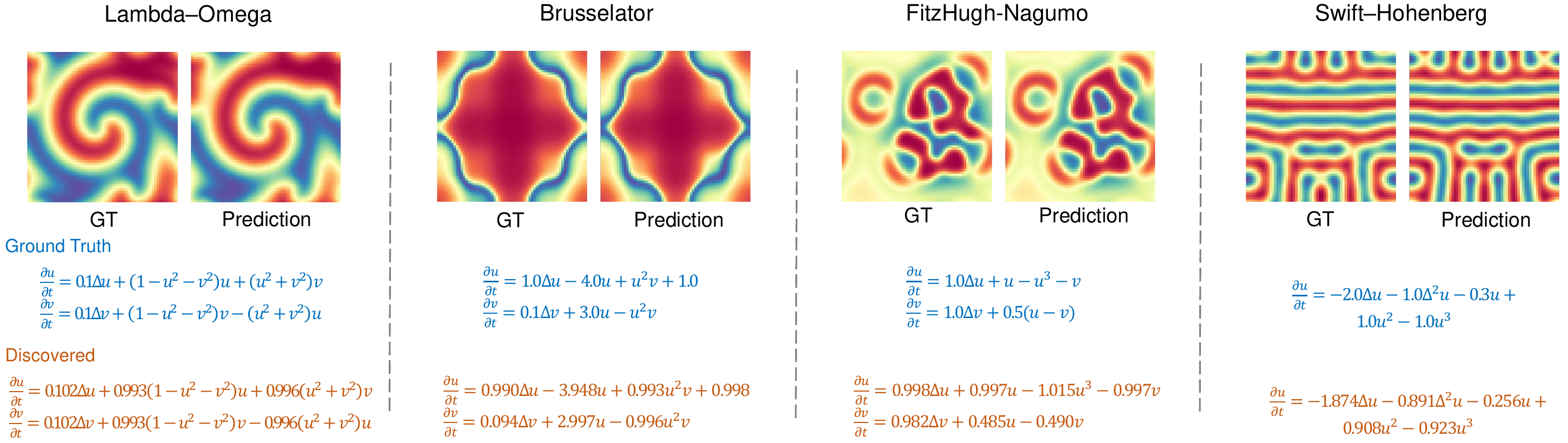}
    \caption{Reasoning results of physical fields: equations in blue is ground truth; equations in orange is inferred by PixelsPhys.}
    \label{fig:pde}
    \vspace{-0.25cm}
\end{figure*}

\subsection{Discovery of Object-level Dynamics}

Consistent with prior works~\cite{luan2022distilling,udrescu2021symbolic, jadhav2022dominant}, we utilize videos of object motion to assess the model's ability to accurately derive symbolic motion equations from high-dimensional observations.


\subsubsection{Datasets}
We validate on five dynamical systems (Figure~\ref{fig:motion}): \textit{Linear}, \textit{Cubic}, \textit{Circular}, \textit{Van Der Pol (VDP)}, and \textit{Glider}. These equations include various characteristics such as linear terms, nonlinear terms, and significant differences in time scales~\citep{guckenheimer1980dynamics}, which together form an evaluation benchmark across different levels of difficulty. 
Video generation details are provided in Appendix~\ref{app:data}.


\subsubsection{Setups}
We categorize the baselines into two classes based on interpretability. The first class encodes latent vectors and learns the implicit dynamics, including AE-SINDy~\cite{champion2019data} and Latent-ODE~\cite{chen2018neural}. The second class explicitly predicts object coordinates to infer their governing equations, represented by Coord-Equ~\cite{luan2022distilling}. 
We train on the initial 200 steps and evaluate using the $R^2$ score on 1,000-step extrapolated coordinate trajectories. 
For AE-SINDy and Latent-ODE, we align their latent predictions to the ground-truth coordinates via Procrustes analysis before evaluation (Appendix~\ref{app:Procrustes}).
Additionally, for methods capable of inferring symbolic expressions, we evaluate two equation-related metrics: 1) Terms Found: A binary score (Yes/No) indicating if all true terms are correctly identified. 2) False Positives: The number of incorrect terms included in the final equation, measuring parsimony.


\subsubsection{Main Results}
Table~\ref{tab:motion} and Figure~\ref{fig:motion} reveal three key insights.
First, implicit methods (Latent-ODE and AE-SINDy) suffer from extrapolation collapse ($R^2 \approx 0$). This empirically validates the necessity of our multi-granularity design, demonstrating that generic representations fail to capture rigid-body physics without explicit object-level parsing.
Second, Pixel2Phys demonstrates superior parsimony and robustness. Compared to Coord-Equ, our framework significantly reduces false positives and achieves near-perfect long-term prediction, with all discovered symbolic expressions detailed in Appendix~\ref{app:add_results}.
Finally, the \textit{Glider} case highlights Pixel2Phys's practicality: although the exact symbolic matching is not achieved due to complex trigonometric terms, the near-perfect extrapolation ($R^2=0.9995$) and visual overlap in Figure~\ref{fig:motion}e confirm that Pixel2Phys successfully captured the underlying physical attractor.
Furthermore, Pixel2Phys transcends simulated settings, successfully recovering gravitational laws from real-world videos, despite complex backgrounds and noisy tracking (see Appendix~\ref{app:real_cases}).

\subsection{Discovery of Pixel-level Dynamics}
In these scenarios, videos represent discrete grid samplings of time-varying fields driven by PDEs~\cite{rudy2017data,champion2019data}. The objective is to capture these pixel-level interactions to derive the underlying PDE equations.

\subsubsection{Datasets}
We conduct experiments on four representative reaction-diffusion equations: Lambda–Omega (LO)~\citep{champion2019data}, Brusselator (Bruss)~\citep{lopez2022gd}, FitzHugh–Nagumo (FHN)~\citep{vlachas2022multiscale} and Swift–Hohenberg (SH)~\citep{swift1977hydrodynamic}.
We solve the differential equations numerically as datasets. The details of the data generation are provided in Appendix~\ref{app:data}.

\subsubsection{Setups}
Baselines are categorized into black-box neural operators and symbolic regression methods. 
Models predict evolution over 1,000 steps (200 for SH system) from initial frames. We evaluate performance using root mean square error (RMSE) and valid prediction steps (VPS), defined as the duration where prediction error remains below a specific threshold (details in Appendix~\ref{app:baselines}).


\subsubsection{Main Results}
Table~\ref{tab:pde} reveals that black-box neural operators suffer from severe error accumulation in long-term rollout and result in extremely low valid prediction steps, confirming that implicit approximations fail to maintain dynamical stability without physical constraints. 
Existing symbolic baselines also show significant fragility as SGA-PDE and LLM-PDE frequently fail to converge or yield suboptimal fits marked as NaN. 
SGA tends to overfit due to the unconstrained search space of genetic algorithms while LLM-PDE lacks visual perception and biases towards over-simplified expressions that miss accurate terms as detailed in the full equation list in Appendix~\ref{app:add_results}. 
In contrast, Pixel2Phys consistently achieves the lowest RMSE and near-perfect stability across all datasets by integrating precise numerical tools within a reasoning loop. Figure~\ref{fig:pde} further demonstrates that our framework correctly identifies complex high-order operators like the bi-harmonic term to capture the exact governing mechanism.
Moreover, our framework scales to complex real-world PIV datasets, accurately recovering 2D Navier-Stokes components (detailed in Appendix~\ref{app:real_cases}).

\subsection{Discovery of Representation-level Dynamics}
This category involves real-world scientific recordings, which suffer from low signal-to-noise ratios due to uncontrolled lighting and sensor noise. 
Consequently, effective physical components are embedded implicitly. The goal is to discover a compact representation space that filters visual noise to capture these implicit evolution mechanisms.

\subsubsection{Datasets}
We collect six videos: four visualizing Kármán vortex streets (fluid dynamics)~\cite{noack2003hierarchy,taira2017modal} and two recording Belousov-Zhabotinsky reactions (chemical oscillators)~\cite{hudson1981chaos}. 
Both represent canonical complex systems governed by low-dimensional attractors but manifested through high-dimensional, noisy visual patterns. 
All videos are cropped and converted to grayscale (details in Appendix~\ref{app:data}).


\subsubsection{Setups}
We benchmark against FNO, Latent-ODE, and the advanced video generation model Wan2.2~\cite{wan2025wan}. 
Given the limited sequence length (fewer than 300 frames), models are trained on the full sequence and evaluated on their ability to autoregressively reconstruct the entire video from the first frame. 
For Wan2.2, we freeze pretrained weights and use GPT-4o to generate descriptive text prompts for conditioning.
In addition to RMSE, we also used vorticity error~\cite{jeong1995identification} to evaluate the accuracy of vorticity, whose formula is shown in Appendix~\ref{app:metrics}.

\subsubsection{Main Results}
\begin{figure}[t]
    \centering
    \begin{subfigure}{0.43\linewidth}
        \centering
        \includegraphics[width=\linewidth]{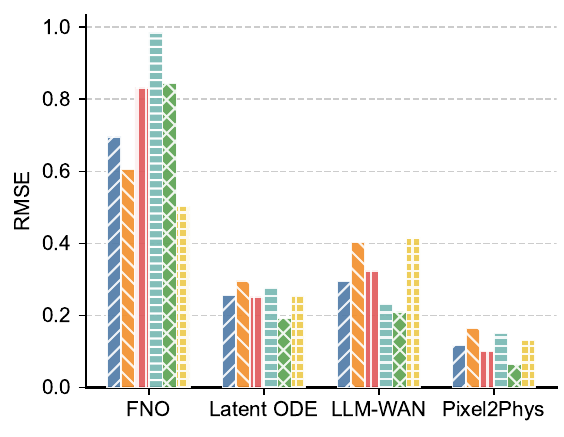}
        \caption{RMSE}
    \end{subfigure}
    \hfill
    \begin{subfigure}{0.56\linewidth}
        \centering
        \includegraphics[width=\linewidth]{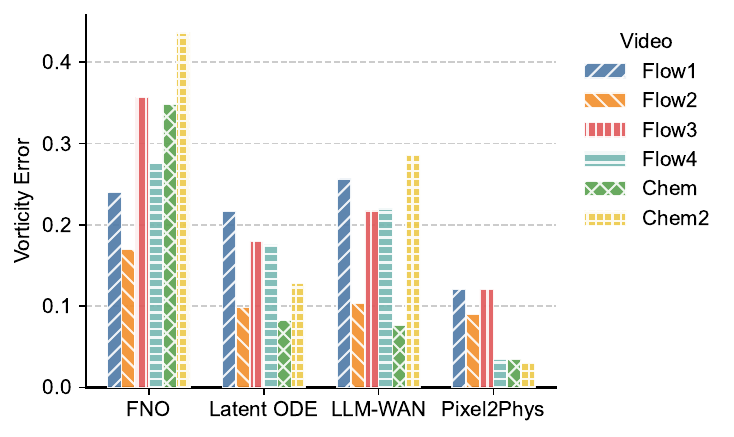}
        \caption{Vorticity Error}
    \end{subfigure}
    \caption{Comparison of prediction errors across all models on physical phenomenon videos.}
    \label{fig:video}
    \vspace{-0.2cm}
\end{figure}

\begin{figure}[t]
    \centering
    \includegraphics[width=\linewidth]{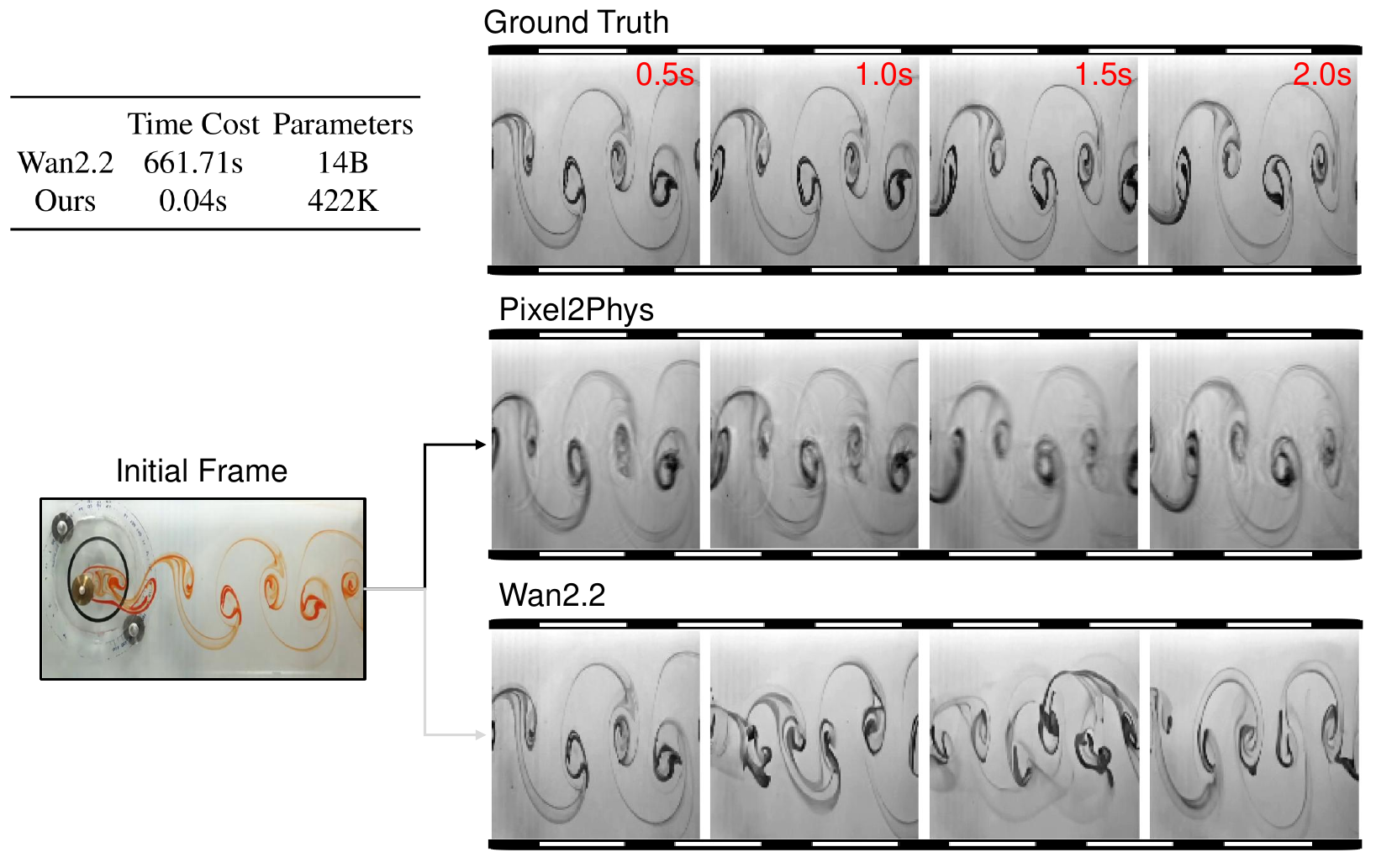}
    \caption{Prediction results of PixelsPhys and Wan2.2 for the Water Flow video.}
    \label{fig:video_flow1}
    \vspace{-0.3cm}
\end{figure}

Figure~\ref{fig:video_flow1} presents a compelling comparison where the 14B-parameter Wan2.2 generates visually realistic textures, it fails to maintain dynamical consistency, evidenced by the spatial drift of vortices at $1.5s$ and $2.0s$ relative to the ground truth. 
This stems from the qualitative nature of generative prompts (see Appendix~\ref{app:prompts}), whereas Pixel2Phys distills quantitative governing laws (e.g., eigenvalues $\lambda_{2,3} \approx \pm 2.136i$ in Figure~\ref{fig:intro}c) that dictate a strict oscillation period ($T \approx 0.294$) to ensure precise alignment. 
Visually, our predictions appear less textured, which is attributed to the selective filtering of the co-optimization mechanism. Irrelevant components like uneven lighting are discarded to extract pure dynamics on the intrinsic manifold. This physical fidelity is further corroborated by the quantitative results in Figure~\ref{fig:video}, where Pixel2Phys consistently achieves the lowest prediction error, confirming that the discovered parsimonious law successfully captures the dominant high-dimensional behavior despite the filtration of visual noise.

\subsection{Ablation Study and Robustness}
To verify the necessity of co-optimization, we replace the Plan Agent with a static serial workflow. 
As detailed in Appendix~\ref{app:ablation}, the absence of the equation feedback loop results in a highly rugged variable space, failing to distill parsimonious laws. 
We further test robustness by substituting the LLM backbone with smaller-scale models. Results in Appendix~\ref{app:robustness} show that Pixel2Phys maintains superior accuracy even with limited reasoning capacity, demonstrating that our collaborative agentic architecture effectively reduces the dependence on raw model scale. Finally, detailed case studies visualizing the step-by-step reasoning process for each video category are provided in Appendix~\ref{app:case_study}.

\section{Conclusion}

In this work, we present Pixel2Phys, a framework that automates the discovery of governing laws from visual dynamics. 
By coordinating specialized agents, our approach replaces static pipelines with an iterative co-optimization process. Crucially, it utilizes preliminary symbolic laws to reversely guide visual variable extraction, effectively resolving the coupling between variable extraction and law discovery.
Experiments show that Pixel2Phys recovers parsimonious equations and achieves robust long-term extrapolation, marking a solid step towards interpretable visual models.
{
    \small
    \bibliographystyle{ieeenat_fullname}
    \bibliography{reference}
}


\end{document}